# Column-Oriented Storage Techniques for MapReduce

Avrilia Floratou
University of
Wisconsin–Madison

Jignesh M. Patel
University of
Wisconsin–Madison

Eugene J. Shekita
IBM Almaden
Research Center

Sandeep Tata
IBM Almaden
Research Center

## ABSTRACT

Users of MapReduce often run into performance problems when they scale up their workloads. Many of the problems they encounter can be overcome by applying techniques learned from over three decades of research on parallel DBMSs. However, translating these techniques to a MapReduce implementation such as Hadoop presents unique challenges that can lead to new design choices. This paper describes how column-oriented storage techniques can be incorporated in Hadoop in a way that preserves its popular programming APIs.

We show that simply using binary storage formats in Hadoop can provide a 3x performance boost over the naive use of text files. We then introduce a column-oriented storage format that is compatible with the replication and scheduling constraints of Hadoop and show that it can speed up MapReduce jobs on real workloads by an order of magnitude. We also show that dealing with complex column types such as arrays, maps, and nested records, which are common in MapReduce jobs, can incur significant CPU overhead. Finally, we introduce a novel skip list column format and lazy record construction strategy that avoids deserializing unwanted records to provide an additional 1.5x performance boost. Experiments on a real intranet crawl are used to show that our column-oriented storage techniques can improve the performance of the map phase in Hadoop by as much as two orders of magnitude.

## 1. INTRODUCTION

Over the last few years, there has been tremendous growth in the need for large scale data processing systems. These systems were once the province of parallel database management systems (DBMSs), but lately the MapReduce paradigm has gained substantial momentum. There is currently an on-going tussle between proponents of these two paradigms [17, 18, 28], with each side claiming strong advantages over the other. However, there is a growing sense that there are advantages to both paradigms, and that techniques which have been successfully used in one can be used to fix deficiencies in the other. Performance is one area in particular where parallel DBMSs currently enjoy an advantage over MapReduce [28].

Hadoop [2] is the popular open-source implementation of MapReduce. In this paper, we describe how the column-oriented storage techniques found in many parallel DBMSs can be used to dramatically improve Hadoop's performance. Our work is motivated by observing the needs of real corporate Hadoop users. These users are familiar with parallel DBMS technology, but they still pick Hadoop for certain applications because of its ease of use, low cost scaling, fault tolerance on commodity hardware, and programming flexibility.

Corporate users tend to be thrilled by how quickly they can get things working in Hadoop. However, as they try to scale up their workloads, they often face a pain point with respect to performance. For example, one user we worked with at a large consumer bank was trying to use Hadoop to process the logs from web applications. They started with raw log files in text format for a single web application. As logs from additional applications were added and the retention period for the logs grew to 90 days, the 20-node Hadoop cluster that they started with could no longer generate reports in a reasonable amount of time. Critics of MapReduce would argue that such users would be better off with a parallel DBMS, but given Hadoop's current advantages for certain applications, and the growing investment in Hadoop by many users, this is not a realistic option. The goal is to fix Hadoop, not replace it with a parallel DBMS.

We have observed a recurring pattern of performance issues in Hadoop that are related to: (a) the use of complex data types such as arrays, maps, and nested records, which are common in many MapReduce jobs, (b) the ability to write arbitrary map and reduce functions in a programming language instead of using a declarative query language, and (c) Hadoop's choice of Java as its default programming language. These issues do not appear in a parallel DBMS; they are unique to the MapReduce paradigm and Hadoop in particular. The column-oriented storage techniques we describe are specifically designed to address these issues.

Besides column-oriented storage techniques, it should be clear that other DBMS techniques can also be leveraged to improve Hadoop's performance, such as efficient join algorithms and indexing [19, 23, 22]. These techniques are beyond the scope of this paper but should be complementary to the ones described here.

## 1.1 Our Contributions

We first present the design and implementation of a column-oriented, binary storage format for Hadoop. We describe how such a format interacts with the replication policy of the Hadoop Distributed File System (HDFS) and the necessary mechanisms to co-locate column data. We demonstrate through experiments that Hadoop can leverage this storage format without incurring a large penalty for reconstructing records from the constituent columns. Such a storage format assumes that the application is willing to pay a one-time loading cost to organize the input data in the appropriate column-oriented fashion. As argued in earlier papers [28], this is a reasonable assumption to make for datasets that are expected to be analyzed multiple times.

The use of text storage formats rather than binary storage formats in performance evaluations of MapReduce has been criticized [18] but never quantified. Recent evaluations of Hadoop [12, 28] used text formats, so it has not been clear how much Hadoop can actually benefit from binary formats. We show that simply switching to a binary storage format can improve Hadoop's scan performance by 3x.

We identify performance challenges specific to complex data types in Hadoop, and describe a novel skip list column format that enables lazy record construction. The lazy record construction we describe is inspired by the late materialization techniques used in column-oriented DBMSs [11]. We also examine techniques that allow lazy decompression in Hadoop. We show that compression techniques like LZO [5] may be too CPU intensive for many MapReduce jobs. Experiments on a real dataset show that lightweight dictionary compression schemes, which provide worse compression ratios than LZO but incur lower CPU overhead during decompression, may be a better alternative for complex data types. We show that these methods can result in speedups of up to 1.5x over an eager record construction strategy.

It is important to emphasize that our column-oriented techniques leverage extensibility features that are already in Hadoop, so no modifications to the core of Hadoop are required. Moreover, our techniques do not require the use of a declarative query language and are designed to work with hand-coded MapReduce jobs. Applications using popular serialization frameworks like Avro [1], Thrift [8], or Protobufs [7] can benefit from our techniques with almost no modification. In aggregate, our techniques can improve the performance of the map phase of a Hadoop job by as much as two orders of magnitude, and the overall job by over an order of magnitude.

## 2. HADOOP BACKGROUND

We first provide some background on Hadoop along with a description of the extensibility points that were used to implement our column-oriented storage format.

Consider the example MapReduce job shown in Figure 1. This is a simplified version of a real Hadoop job that analyzes a collection of crawled documents and finds all the distinct "content-types" reported for URLs that contain the pattern "ibm.com/jp". We initially focus on the main program where the job is configured with an `InputFormat` and `OutputFormat`.

An `InputFormat` is an important abstraction and extensibility point in Hadoop. It is responsible for two main functions:

```
class MyMapper {
  void map (NullWritable key, Record rec) {
    String url = (String) rec.get("url");
    if (url.contains("ibm.com/jp"))
      output.collect(null,
        rec.get("metadata").get("content-type"));
  }
}

class MyReducer {
  void reduce(NullWritable key, Iterator<Text> vals){
    HashSet<Text> distinctVals = new HashSet<Text>();
    for (Text t: vals)
      distinctVals.add(t);
    for (Text t: distinctVals)
     output.collect(null, key);
  }
}

main() {
  Job job = new Job();
  job.setMapperClass(MyMapper);
  job.setReducerClass(MyReducer);
  job.setInputFormat(SequenceFileInputFormat.class);
  SequenceFileInputFormat.addInputPath(job,"/data/jan");
  job.setOutputFormat(TextOutputFormat.class);
  TextOutputFormat.setOutputPath("/output/job1");
  JobRunner.submit(job);
}
```

**Figure 1: Example MapReduce job.**

first, to generate `splits`[1] of the data that can each be assigned to a map task; and second, to transform data on disk to the typed key and value pairs that are required by the map function. An `InputFormat` implements the following three methods:

**addInputPath()** is used to specify input data sets when configuring the job.

**getSplits()** is used by the Hadoop scheduler to get a list of splits for the job.

**getRecordReader()** is invoked by Hadoop to obtain an implementation of a `RecordReader`, which is used to read the key and value pairs from a given split.

Hadoop provides different `InputFormats` to consume data from text files, comma separated files, etc. For example, the MapReduce job in Figure 1 is configured to use a `SequenceFileInputFormat`. A `SequenceFile` stores key and value pairs in a standard, serialized binary format. The dual of an `InputFormat` in Hadoop is an `OutputFormat`, which is responsible for transforming the key-value pairs output by a MapReduce job to a disk format.

The key and value pairs consumed by map and reduce functions can be of any object type. In this paper, we assume that `Record` objects are used for values. Serialization frameworks like Avro [1], Protocol Buffers [7], or Thrift [8] can be used to provide a record abstraction with methods to convert records to raw bytes, read them from disk, or pass them between map and reduce tasks. The attributes of a record are accessed using a Java `get(name)` method, which takes the name of an attribute as a parameter. Type casting is usually required to access an attribute. A more detailed discussion of the Avro record abstraction used in our experiments can be found in Appendix A.

---
[1]A split is the unit of scheduling and is a non-overlapping partition of the input data that is assigned to a map task.

```
URLInfo {
  Utf8 url,
  Utf8 srcUrl,
  time fetchTime,
  Utf8[] inlink,
  Map<String, String> metadata,
  Map<String,String> annotations,
  byte[] content
}
```

Figure 2: Example schema with complex types.

## 3. CHALLENGES

This section discusses the performance challenges outlined in the introduction in more detail.

### 3.1 Complex Data Types

In large scale data analysis, it is often convenient to use complex types like arrays, maps, and nested records to model data. Recent studies [16, 25] have argued that it is better to use native, nested representations of complex types in read-mostly workloads. This is in contrast to flattening complex types into normalized relational tables.

Figure 2 shows an example schema with complex types. The example is taken from an actual intranet search application where documents are crawled and stored along with metadata, extracted annotations, and inlinks. The annotations and the metadata vary widely from page to page and are therefore stored using maps. Inlinks are stored in an array. The use of complex types causes two major problems: deserialization costs and the lack of effective column-oriented compression techniques.

### 3.2 Serialization and Deserialization

Serialization is the process of converting a data structure in memory into bytes that can be transmitted over the network or written out to disk. Deserialization is the inverse of this process.

The overhead of deserializing and creating the objects corresponding to a complex type can be substantial. Previous studies [16, 23] have noted the importance of paying attention to the cost of deserialization and object creation in Hadoop. Most column-oriented DBMSs are implemented in C++, allowing column data from disk to be directly accessed in memory as an array without any deserialization overhead [9]. For example, suppose we want to compute the sum of 1 million integers in a file. In C++, the integers can be read into a memory buffer, and an array pointer can be cast to the beginning of the buffer. Then the array's elements can be summed directly in a tight loop. Java, on the other hand, would require deserializing each integer from the memory buffer before summing it.

We conducted an experiment to measure this overhead. Our experiments showed that the CPU overhead of deserialization and object creation is so significant that it can quickly become a bottleneck in Hadoop. This overhead even affects simpler data types such as integers. For more details on this experiment, see Appendix B.

### 3.3 Compression

Using compression results in lower I/O costs at the expense of higher CPU costs. In general, column-oriented storage formats tend to exhibit better compression ratios since data within a column tends to be more similar than data across columns. Previous studies have looked at com-pression in the context of a column-oriented DBMS [10]. However complex types, which are not as amenable to techniques like run-length compression, dictionary compression, offset encoding, etc., were not considered in those studies.

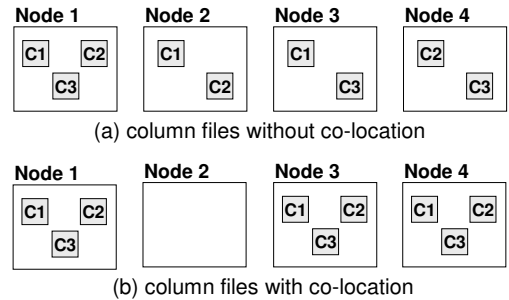

Figure 3: Co-locating column files.

Lightweight compression schemes are critical in Hadoop. Schemes like ZLIB, which achieve excellent compression ratios, incur substantial CPU overhead during decompression. As a result, LZO [5] is commonly used in Hadoop to provide reasonable compression ratios with low decompression overhead. In Section 5.3, we describe a lightweight dictionary compression scheme for our column-oriented storage format that works well with complex types and provides better performance than LZO and ZLIB.

### 3.4 Query Language vs. Programming API

In contrast to a DBMS, where a declarative query language is compiled into a set of runtime operators, the basic MapReduce framework provides only a programming API. Unfortunately, many of the advanced techniques used by column-oriented DBMSs are not feasible with hand-coded map and reduce functions – the programming task would be too difficult for a human. These techniques include operating on compressed data [10], the use of SIMD instructions [14], and late materialization [11]. This paper focuses on hand-coded MapReduce jobs written against the programming API of Hadoop. The techniques described here are also applicable to declarative languages on Hadoop such as Pig [27], Hive [3], or Jaql [4].

## 4. COLUMN-ORIENTED STORAGE

We now describe the design and implementation of our column-oriented storage format and its interaction with Hadoop's data replication and scheduling.

### 4.1 Replication and Co-location

A straightforward way to implement a column-oriented storage format in Hadoop is to store each column of the dataset in a separate file. This imposes two problems. First, how can we generate roughly equal sized splits so that a job can be effectively parallelized over the cluster? Second, how do we make sure that the corresponding values from different columns in the dataset are co-located on the same node running the map task?

The first problem can be solved by horizontally partitioning the dataset and storing each partition in a separate subdirectory. Each such subdirectory now serves as a split. The second problem is harder to solve. HDFS uses 3-way block-level replication to provide fault tolerance on commodity servers, but the default block placement policy does not provide any co-location guarantees.

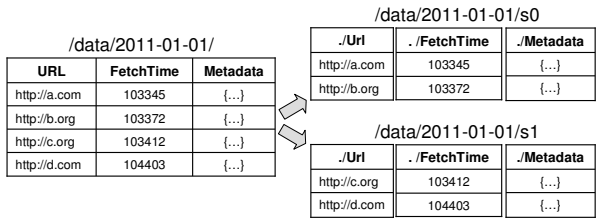

**Figure 4: Data layout with COF.**

Consider a dataset with three columns C1, C2, and C3. Assume the columns are stored in three different files, and for simplicity, also assume that each file occupies a single HDFS block. In practice, a large file could span many HDFS blocks. The files of C1-C3 need to be accessed together as a split, but with Hadoop's default placement policy, they could be randomly spread over the cluster. Figure 3a illustrates what can happen with Hadoop's default placement policy. C1-C3 are co-located on Node 1 but not co-located on any other node. Suppose a map task is scheduled for the split consisting of C1-C3 but Node 1 is busy. In that case, Hadoop would schedule the map task on some other node, say Node 2, but performance would suffer, since C3 would have to be remotely accessed.

Recent work on the RCFile format [20] avoids these problems by resorting to a PAX [13] format instead of a true column-oriented format. RCFile takes the approach of packing each HDFS block with chunks called *row-groups*. Each row-group contains a special synchronization marker at the start, followed by a metadata region, and then a data region, with the data region laid out in a column-oriented fashion. The metadata describes the columns in the data region and their starting offsets, as well as the number of rows in the data region. Since all the columns are packed into a single row-group, and each row-group can function independently as a split, it avoids the two challenges that arise when storing columns separately.

While RCFile is simple and fits well within Hadoop's constraints, it has a few drawbacks. Since the columns are all interleaved in a single HDFS block, efficient I/O elimination becomes difficult because of prefetching by HDFS and the local filesystem. Tuning the row-group size and the I/O transfer size correctly also becomes critical. With larger I/O transfer sizes like 1MB, records that contain more than four columns show very poor I/O elimination characteristics with the default RCFile settings. Finally, extra metadata needs to be written for each row group, leading to additional space overhead.

The next section describes an alternative to RCFile that uses separate files for each column and still avoids these problems. Experiments in Section 6 will show that this new format can significantly outperform RCFile.

### 4.2 The CIF Storage Format

We solved the problem of co-locating associated column files by implementing a new HDFS block placement policy. HDFS allows its placement policy to be changing by setting the configuration property "dfs.block.replicator.classname" to point to the appropriate class. This feature has been present since Hadoop 0.21.0 and does not require recompiling Hadoop or HDFS.

`ColumnPlacementPolicy` (CPP) is the class name of our column-oriented block placement policy. For simplicity, we will assume that each column file occupies a single HDFS block and describe CPP as though it works at the file level. In effect, CPP guarantees that the files corresponding to the different columns of a split are always co-located across replicas. Figure 3b shows how C1-C3 would be co-located across replicas using CPP. Subdirectories that store splits need to follow a specific naming convention for CPP to work. Files that do not follow this naming convention, are replicated using the default placement policy of HDFS.

We implemented the logic for our column-oriented storage format in two classes: the `ColumnInputFormat` (CIF) and the `ColumnOutputFormat` (COF). Data may arrive into Hadoop in any format. Once it is in HDFS, a parallel loader is used to load the data using COF.

Consider an example scenario involving the data described in Figure 2. Assume that crawled data arrives at regular intervals and that a day's worth of data has arrived and needs to be stored in "/data/2011-01-01". When a dataset is loaded into a subdirectory using COF, it breaks the dataset into smaller horizontal partitions. Each partition, referred to as a *split-directory*, is a subdirectory with a set of files, one per column in the dataset. An additional file describing the schema is also kept in each split-directory. Figure 4 shows the layout of data using COF, with split-directories s0 and s1.

When reading a dataset, CIF can actually assign one or more split-directories to a single split. The column files of a split-directory are scanned sequentially and the records are reassembled using values from corresponding positions in the files. Projections can be pushed into CIF by supplying it with a list of columns. This can be done while configuring a MapReduce job as follows:

`ColumnInputFormat.setColumns(job, "url, metadata");`

The record objects created by CIF are populated only with the fields that are selected. The files corresponding to the remaining columns are not scanned.

### 4.3 Discussion

A major advantage of CIF over RCFile is that adding a column to a dataset is not an expensive operation. This can be done by simply placing an additional file for the new column in each of the split-directories. With RCFile, adding a new column is a very expensive operation – the entire dataset has to be read and each block re-written.

Adding columns is well known to be an important feature. This is a particularly common operation when the dataset needs to be augmented with derived columns computed from the existing columns. We have also seen the need for this feature when a customer starts by extracting a set of columns from raw input files (such as logs) into organized storage for efficient querying. As business needs evolve, additional columns from the raw input files need to be moved to organized storage.

Experiments in Section 6 will show that CIF does not pay a performance penalty for this flexibility advantage over RCFile. In fact, CIF overcomes some drawbacks of RCFile with respect to metadata overheads, poor prefetching, and I/O elimination.

On the other hand, a potential disadvantage of CIF is that the available parallelism may be limited for smaller datasets. Maximum parallelism is achieved for a MapReduce job when the number of splits is at least equal to the number of map slots, say $m$. RCFile allows fine grained splits at the row-group level (4MB) when compared to split-directories in CIF (typically 64 MB). For RCFile, assuming that each HDFS block has $r$ row-groups, maximum parallelism is available when the total dataset size is greater than $m/r$ blocks. With CIF, this happens when there are at least $m$ split-directories. If we choose split-directories containing $c$ blocks worth of data in each directory (where $c$ is the number of columns), full parallelism is available when the dataset size exceeds $m \times c$ blocks.

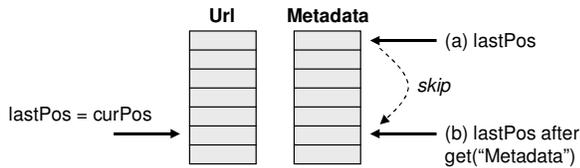

Figure 5: Lazy record construction.

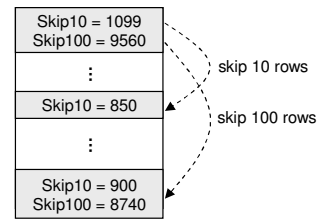

Figure 6: Skip list format for complex types.

Assuming a typical cluster with 200 map slots and 64M blocks, a dataset with 10 columns would need to be at least 128GB in size before full parallelism is reached. Since we expect to deal with datasets in the terabyte range on Hadoop, we expect to be able to utilize all the available parallelism in a cluster with CIF. In practice even with RCFile, large row-groups are preferred since they minimize metadata overhead and improve I/O elimination (see Figure 9 in Appendix B).

Load balancing with CIF and CPP happens at a coarser granularity (per split-directory) using the *same* algorithms as the default placement policy. This is because CPP chooses the location of the first block of a split-directory using the default placement policy. All the remaining blocks in the split directory are then placed on the same set of nodes.

In summary, CIF offers flexibility and some performance benefits over RCFile. This advantage comes at the cost of needing to install a special block-placement policy for HDFS and potentially limiting the amount of parallelism for smaller datasets. We do not expect either of these considerations to be a problem for large deployments. A deeper analysis of load-balancing and re-replication after failures are important avenues for future work.

## 5. LAZY RECORD CONSTRUCTION

In this section, we describe our lazy record construction technique, which is used to mitigate the deserialization overhead in Hadoop, as well as eliminate disk I/O. The basic idea behind lazy record construction is to deserialize only those columns of a record that are actually accessed in a map function. Consider the example MapReduce job in Figure 1 that was described earlier. The metadata column is accessed in the map function only for records where the URL column contains the pattern "ibm.com/jp". Using lazy record construction, we can avoid deserializing the metadata column for the records where the URL column does not contain this pattern.

### 5.1 Implementation

CIF can be configured to use one of two classes for materializing records, namely, `EagerRecord` or `LazyRecord`. Both of these classes implement the same `Record` interface. As a result, the map function code looks the same, regardless of which class is instantiated.

`EagerRecord` eagerly deserializes all the columns that are being scanned by CIF. `LazyRecord` is slightly more complicated. Internally, `LazyRecord` maintains a split-level *curPos* pointer, which keeps track of the current record the map function is working on in a split. It also maintains a *lastPos* pointer per column file, which keeps track of the last record that was actually read and deserialized for a particular column file. Both pointers are initialized to the first record of the split at the start of processing.

Each time `RecordReader` is asked to read the next record, it increments *curPos*. No bytes are actually read or deserialized until one of the `get()` methods is called on the resulting `Record` object. Consider the example in Figure 5. Since get("url") is called on every record, *lastPos* is always equal to *curPos* for the URL column. However, for the metadata column, *lastPos* may lag behind *curPos* if there are records where the URL column does not contain the pattern "ibm.com/jp". When the URL column contains this pattern and get("metadata") is called, *lastPos* skips ahead to *curPos* before the metadata column is deserialized.

Note that complex column types with variable lengths are the main reason both the split-level *curPos* and per column file *lastPos* pointers are needed for lazy record construction. Ostensibly, it might seem like just a *curPos* pointer per column file could be used without a *lastPos* pointer. However, in that case, each next record call would require all the columns to be deserialized to extract length information to update their respective *curPos* pointers. This in turn would defeat the purpose of lazy record construction.

### 5.2 Skip List Format

A skip list format [29] can be used within each column file to efficiently skip records. Figure 6 shows the format used in CIF. A column file contains two kinds of values, regular serialized values and *skip blocks*. Skip blocks contain information about byte offsets to enable skipping the next N records, where N is typically configured for 10, 100, and 1000 record skips.

Column files support a `skip()` method that is called by `LazyRecord` as skip(*curPos* - *lastPos*). If a column file is not formatted as as a skip list, each record is skipped individually, resulting in no deserialization or I/O savings. The cost for creating a skip list format is paid once at load time. Experiments in Appendix B show that the additional overhead incurred during loading is minimal.

### 5.3 Compression

We propose two schemes to compress columns of complex data types: compressed blocks, and dictionary compressed skip lists. Section 6 compares these two schemes. Both schemes are amenable to lazy decompression where portions of the data that are not accessed in the map function are not decompressed.

**Compressed Blocks:** This scheme uses a standard compression algorithm to compress a block of contiguous column values. Multiple compressed blocks may fit into a single HDFS block. The compressed block size is set at load time. It affects both the compression ratio and the decompression overhead. A header indicates the number of records in a compressed block and the block's size. This allows the block to be skipped if no values are accessed in it. However, when a value in the block is accessed, the entire block needs to be decompressed. LZO is generally chosen for the compression algorithm in favor of other strategies like ZLIB when low decompression overhead is more important than the compression ratio. We study both LZO and ZLIB in Section 6.

**Dictionary Compressed Skip List:** This scheme is tailored for map-typed columns. It takes advantage of the fact that the keys used in maps are often strings that are drawn from a limited universe. Such strings are well suited for dictionary compression. We build a dictionary of keys for each block of map values and store the compressed keys in a map using a skip list format. This scheme often provides a worse compression ratio than LZO but compensates with lower CPU overhead for decompression. The main advantage of this scheme is that a value can be accessed without having to decompress an entire block of values.

## 6. EXPERIMENTS

In this section, we present experimental results demonstrating that column-oriented storage techniques can be effectively used in Hadoop. We compare CIF with popular formats in use, namely text files (TXT), `SequenceFiles` (SEQ), and RCFile.

### 6.1 Experimental Setup

The experiments were run on a cluster with 42 nodes connected by a 1Gbit ethernet switch. Two nodes were reserved to run the Hadoop jobtracker and the namenode. The remaining 40 nodes were used for HDFS and MapReduce. Each node had 8 cores (via two quad-core 2.4Ghz sockets), 32GB of main memory, and five locally attached 500GB SATA 1.0 disks. Datanodes spread their data across four of these disks. Hadoop version 0.21.0 was used, and was configured to run 6 mappers per node (i.e., 6 map *slots* per node) and 1 reducer per node.

### 6.2 Benefits of Column-Oriented Storage

The first experiment was a microbenchmark to verify that using CIF can indeed make scans faster compared to using SEQ and TXT. As in [23], these experiments were run on a single node of the cluster. Data was read using standard HDFS and `InputFormat` APIs. We present experiments using the full cluster in Section 6.3.

We used a synthetic dataset generated as follows: Each record consisted of 6 strings, 6 integers, and a map. The integers were randomly assigned values between 1 and 10000. Random strings of length between 20 and 40 were generated over readable ASCII characters. Each map column consisted of 10 items, where the keys were random strings of length 4, and the values were randomly chosen integers. The data was written out in each of the formats. For SEQ, `NullWritables` were used as the keys. A record containing the above fields was used as the value class. The total size of the dataset was 57GB in the SEQ format. The I/O transfer size, `io.file.buffer.size`, was set to 128K. This is a commonly configured value for many deployments. Repeating the experiment with 4KB and 1MB produced similar results and are omitted. The time to scan various projections of the dataset for each of the formats is shown in Figure 7. The filesystem cache was flushed before each experiment. For TXT and SEQ, the time to scan any projection was roughly the same and therefore only one value is reported.

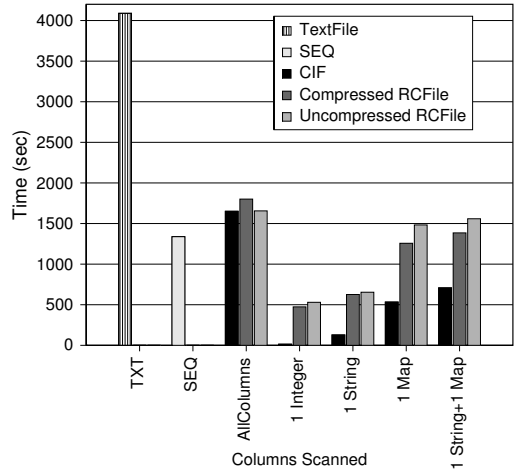

**Figure 7: Microbenchmark comparing Text, SEQ, CIF, and RCFile.**

**Comparison with TXT:** As shown in Figure 7, the scan time with SEQ was approximately 3x faster than TXT. Parsing each line of the text file caused TXT to quickly become CPU-bound, whereas the parsing overhead was avoided in SEQ. This confirms the criticism by Google engineers [18] where they argued that previous studies comparing Hadoop's performance to a parallel DBMS [28, 12] were flawed because of the naive use of text files in Hadoop. Using a binary format like SEQ is a straightforward way to dramatically improve Hadoop's performance.

**Comparison with Sequence Files:** When using CIF, the times for scanning a single integer, string, or map were 2.5x to 95x faster than SEQ. In each case, the speedup is directly attributable to the fact that CIF read much less data than SEQ. When scanning *all* the columns of the dataset, CIF took about 25% longer than SEQ. This is because of the additional seeks that CIF incurred when gathering data from columns stored in different files. In all other cases, CIF is superior to both TXT and SEQ.

**Comparison with RCFile:** The uncompressed RCFile was approximately 69GB and the compressed RCFile was 43GB. The row-group size for RCFile was set to the recommended value of 4MB [20]. When a small number of columns were chosen from the dataset, CIF was more efficient than RCFile at eliminating unnecessary I/O. For the case of a single integer, the worse case for RCFile, CIF was nearly 38x faster than the uncompressed RCFile. Measurements using `iostat` revealed that RCFile read 20x more bytes than CIF even when instructed to scan exactly one column. Some of this overhead is due to the use of inefficient serialization in parts of RCFile. Additionally, it incurred more CPU overhead since it had to interpret the metadata blocks for approximately every 4MB of data.

| Layout | Data Read (GB) | Map Time (sec) | Map Time Ratio | Total Time (sec) | Total Time Ratio |
|---|---|---|---|---|---|
| SEQ-uncomp | 6400 | 1416 | - | 1482 | - |
| SEQ-record | 3008 | 820 | - | 889 | - |
| SEQ-block | 2848 | 806 | - | 886 | - |
| SEQ-custom | 3040 | 754 | 1.0x | 806 | 1.0x |
| RCFile | 1113 | 702 | 1.1x | 761 | 1.1x |
| RCFile-comp | 102 | 202 | 3.7x | 291 | 2.8x |
| CIF-ZLIB | 36 | 12.8 | 59.1x | 77 | 10.4x |
| CIF | 96 | 12.4 | 60.8x | 78 | 10.3x |
| CIF-LZO | 54 | 12.4 | 61.0x | 79 | 10.2x |
| CIF-SL | 75 | 9.2 | 81.9x | 70 | 11.5x |
| CIF-DCSL | 61 | 7.0 | 107.8x | 63 | 12.8x |

**Table 1: Comparison of storage formats, with speedup ratios relative to SEQ-custom.**

When using a compressed RCFile, the running time improved. However, CIF was still faster in all cases. For instance, when a single integer was projected, CIF was 33x faster than the compressed RCFile. For the case where all the columns from the dataset were examined, CIF, compressed RCFile, and the uncompressed RCFile all had approximately the same performance. SEQ was 1.2x faster than the rest. Experiments in Appendix B show that CIF's advantage over RCFile holds for other values of the row-group size.

### 6.3 Comparison of Column Layouts

Next, we compared the performance of different storage formats on a real dataset consisting of crawled pages for an intranet search application. This application currently uses Hadoop for its backend analytics. The data was acquired using the Nutch [6] crawler and stored in HDFS. The schema of the dataset we used is described in Figure 2 and included fields like encoding, language, location, among others.

The MapReduce job used in our experiments found all the distinct content-types reported by web pages from IBM Japan i.e., URLs containing "ibm.com/jp". The content-type for a page was stored in a metadata column with a map data type. The metadata column also included other information returned in the HTTP response for the page. The code of our MapReduce job was very similar to the example in Figure 1. The selectivity of the predicate on the URL was approximately 6%. We executed this job on a 6.4TB subset of the crawl dataset. The total amount of data per node was approximately 160GB.

With SEQ, we tried four variants: uncompressed (SEQ-uncomp), block-compressed (SEQ-block), record compressed (SEQ-record), and a custom format (SEQ-custom), which used an uncompressed sequence file, but compressed the content column using application specific code. We also included the time taken with RCFile with and without Zlib compression enabled (RCFile and RCFile-comp). For CIF, we laid out the metadata column in five different ways: default (CIF), CIF with skip lists (CIF-SL), CIF with block compression using LZO (CIF-LZO) and ZLIB (CIF-ZLIB), and CIF with dictionary compressed skip lists (CIF-DCSL). TXT was omitted from our experiments because of its exceedingly bad performance. `ColumnPlacementPolicy` (CPP) was used for all the CIF experiments.

Table 1 presents the time consumed along with the total bytes read from HDFS for each of the storage formats. As shown, we report the map time and the total time for each storage format. The map time was the average time spent per node in the map phase. This was calculated by taking the total time consumed by all map tasks and dividing it by the number of map slots in the cluster. Measuring the map time allowed us to isolate the improvement offered by different storage formats to just the map phase of our MapReduce job. In contrast, the total time was the wall-clock time for the full MapReduce job to finish.

The results in Table 1 show that the SEQ variants were generally the slowest since they also read the content field, which contains several KB of data for each record. In terms of map time, SEQ-record and SEQ-block were both better than SEQ-uncomp by approximately 1.7x. SEQ-custom was the fastest by a small margin.

The map time of RCFile and RCFile-comp were better than SEQ-custom by 1.1x and 3.7x respectively. Both RCFile and RCFile-comp eliminated some of the I/O for unreferenced columns and read substantially less data than SEQ-custom (3040GB for SEQ-custom vs. 102GB for RCFile-comp).

Turning to the CIF variants, the map time of CIF was 60.8x better than SEQ-custom. This speedup was largely the result of 31.7x less data being read in CIF because of its column-oriented storage format. CIF-ZLIB was slightly worse than CIF, despite reading significantly less data (36GB vs. 96GB). This was because the data reduction in CIF-ZLIB was not enough to compensate for the CPU overhead of its decompression algorithm. The performance of CIF-LZO was similar. We also repeated the experiment with different compression block sizes but did not observe a significant difference.

Continuing down Table 1, the map time of CIF-SL was better then CIF-LZO even though it read more data (75GB vs. 54GB). This was the result of using skip lists and the `LazyRecord` format to avoid deserializing the metadata column unless the URL contained "ibm.com/jp". CIF-DCSL demonstrated the best performance overall, providing a speedup of 107.8x over SEQ-custom. Its decompression algorithm proved to be extremely fast and it also benefited from lazy record construction and the use of skip lists.

The last two columns of Table 1 compare the total time of our MapReduce job with different storage formats. Compared to the map time, the speedup in total time was lower. This is because the total time included the shuffle, sort, and reduce phases of the MapReduce job, which are unaffected by the storage format. However, similar trends were observed, with CIF-DCSL providing the best overall results and a 12.8x speedup over SEQ-custom.

### 6.4 Impact of Co-Location

To measure the impact of co-location, we re-executed the same MapReduce job as above but this time using CIF with the default HDFS block placement policy rather than with CPP. The map time with CPP was 5.1x better than the map time without CPP. CIF with CPP was faster because CPP ensured that no column files had to be remotely accessed.

### 7. RELATED WORK

Pavlo et. al [28] compared the performance of Hadoop with parallel DBMSs and found that Hadoop can perform substantially worse on certain workloads. Subsequent studies have tried to improve on Hadoop using parallel DBMS techniques while still maintaining the flexibility of its APIs. HadoopDB [12] advocated using database nodes to do the actual work and relying on MapReduce only for scheduling and communication. A later study [19] pointed out the drawbacks of HadoopDB and demonstrated, along with other independent efforts [22, 23], how indexing can be incorporated into MapReduce in a less disruptive manner. Using DBMS optimization techniques for declarative query languages like Pig was suggested in [26]. However, our focus is on MapReduce programs written directly in Java. The database community has also been interested in several other aspects of bridging the gap between MapReduce and DBMSs [15].

This paper draws on many of the techniques advocated in the literature for column-oriented DBMSs. The advantage of column-oriented storage for eliminating I/O is well known. However, many of the advanced techniques used in a column-oriented runtime such as careful integration of compression with query execution [10, 24], late materialization [11], use of SIMD instructions [14], and other organizing techniques [21] are challenging to adapt to MapReduce without assuming a declarative query language and a highly specialized runtime for query execution.

A recent paper described Dremel [25], which is a column-oriented storage system used at Google for processing large datasets involving nested types. Dremel shreds nested data into separate columns and reconstructs only the portions needed by a query. Dremel uses a SQL-like language and a special runtime. In contrast to Dremel, we store complex types as a single column and do not shred it into separate columns. In addition to nested records, we also deal with map data types, which are not a focus in Dremel. Our focus is on performance improvement in the context of Hadoop and Java. However, we believe our column-oriented storage techniques complement many of the techniques used in Dremel.

## 8. CONCLUSIONS

The column-oriented storage techniques that have proven so successful in parallel DBMSs can also be used to dramatically improve the performance of MapReduce. However, translating these techniques to a MapReduce implementation such as Hadoop presents unique challenges because of different replication and scheduling constraints, the low-level MapReduce programming API, and the use of complex types in MapReduce jobs. In this paper, we described a new column-oriented binary storage format for Hadoop that is not only compatible with Hadoop's programming APIs but also requires no changes to the core of Hadoop. It includes features such as lightweight compression and lazy record construction to avoid deserializing unwanted records. Experiments on a real intranet crawl were used to show that our column-oriented storage techniques can improve the performance of the map phase in Hadoop by as much as two orders of magnitude, and the overall time of a full MapReduce job by over one order of magnitude.

## 9. ACKNOWLEDGEMENTS

We would like to thank the reviewers of this paper for their constructive comments. This research was supported in part by the National Science Foundation under grant IIS-0963993.

# APPENDIX
## A. RECORD ABSTRACTION

In a MapReduce job, the type of keys and values supplied to the map function depends on the `InputFormat`. The programmer is responsible for supplying a map function that is compatible with the `InputFormat`. For instance, in the case of the job described in Figure 1, the programmer needs to know that the `SequenceFile` being read contains keys of type `NullWritable` and values of type `Record`.

In this paper, we assume that MapReduce jobs are written using a generic class that provides a record abstraction. We use the `Record` interface supplied by the Avro serialization framework. `LazyRecord` and `EagerRecord` described in Section 5 are classes that implement this interface. `ColumnInputFormat` produces keys of type `NullWritable` and values of type `Record`.

Attributes are accessed using a `get(name)` method that takes the name of the attribute as a parameter. The return type for this method is `java.lang.Object`. As a result, type casting is required to access the field values. This `Record` class can be used for records that conform to any schema. Figure 1 illustrates the use of this record abstraction in a map function.

Other serialization frameworks have emerged in the open source that generate a binary representation for a record. Examples include Protocol Buffers [7] and Thrift [8]. These frameworks typically allow developers to specify a schema so that records can be serialized and deserialized efficiently. The schema language supports complex fields like arrays, maps, and nested records.

Avro also supports the notion of a "specific" record. Given a schema, the Avro compiler can be used to produce a Java class containing specific `get` methods for each of the attributes with precise return types. For instance, one could generate a URLInfo class using the schema from Figure 2. The equivalent map function for Figure 1 would be simplified to:

```
map(NullWritable key, URLInfoRecord rec)
{
 if (rec.getUrl().contains.("ibm.com/jp"))
  output.collect(null,
    rec.get("metadata").get("content-type"))
}
```

While we used Avro, the principles we describe in this paper are also applicable to Thrift and Protobufs. A minor advantage of Avro is that code-generation is optional. A generic record class can be used to process data without generating code from a schema. Protobuf and Thrift require code generation by default. Additional work would have been needed to use those frameworks.

Note that as of Version 1.3.3, by default, Avro supplies only a single `get()` method in its generated classes, much like the generic `Record` class. Extending the compiler to generate accessor methods with the appropriate return types is not difficult.

## B. ADDITIONAL EXPERIMENTS
### B.1 Cost of Deserialization

We conducted a simple experiment to illustrate the overhead involved in deserializing simple and complex types. We created a dataset with 1 million records, each 1000 bytes wide. We filled a given fraction $f$ of the 1000 bytes with integers. The remainder of the record was filled with a byte array. The integers require deserialization when the data is scanned. The byte array can be read into the record without any deserialization. We vary the fraction $f$ from 0.0 to 1.0. We measure the time taken to scan this entire dataset in a simple Java program. The data was written to a local file and the filesystem cache was warmed before reading the data. As a result, the entire dataset was present in memory, and no-disk I/O was incurred in any of the cases. We also repeated the experiment in C++. The same single machine configuration that was described in Section 6 was used here.

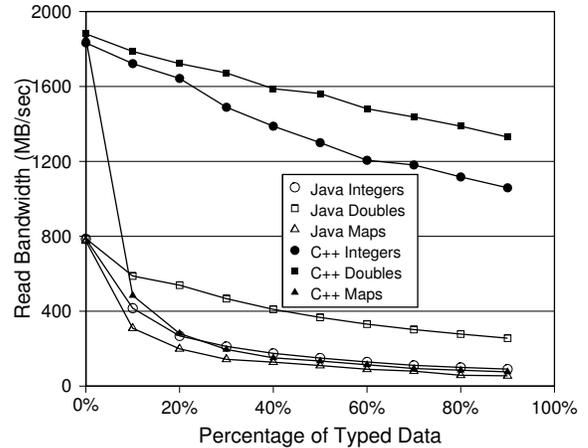

**Figure 8: Microbenchmark examining overhead of serialization and object creation.**

Figure 8 shows the total read bandwidth measured while scanning this dataset as $f$ was varied for different data types: integers, doubles, and maps. For the map implementations, we used `java.util.map` for Java and `std::map` for C++.

As shown, in each case, the read bandwidth drops as $f$ increases. This is because deserializing typed data imposes a larger CPU overhead than reading a simply byte array. As explained in Section 3.2, Java suffers much more from this phenomenon than C++. The read bandwidth of the C++ program is substantially higher for integers and doubles than the Java program.

The case of deserializing maps is more interesting. Each map consisted of 4 entries, the keys were mutable strings and the values were integers. Since maps require new objects to be created, the total overhead of deserializing maps is substantially larger. In fact, as the figure shows, when $f$ exceeds 60%, the rate at which maps are deserialized can be slower than the bandwidth of a typical SATA disk.

### B.2 Tuning RCFile

We studied the impact of varying the row-group size of the RCFile on the scan tests described in Section 6.2. Using the same dataset as before, we varied the row-group sizes as 1MB, 4MB, and 16MB. The running times for scanning various projections are shown in Figure 9.

For the case where a single integer was scanned, CIF read a total of 415MB. RCFile read 16.5GB, 8.5GB, and 4.5 GB for the 1MB, 4MB, and 16MB row-group size settings re-

427

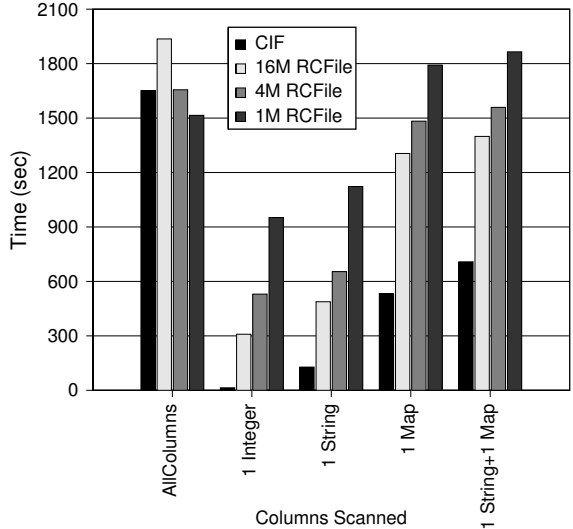

Figure 9: Tuning row-group size for RCFiles

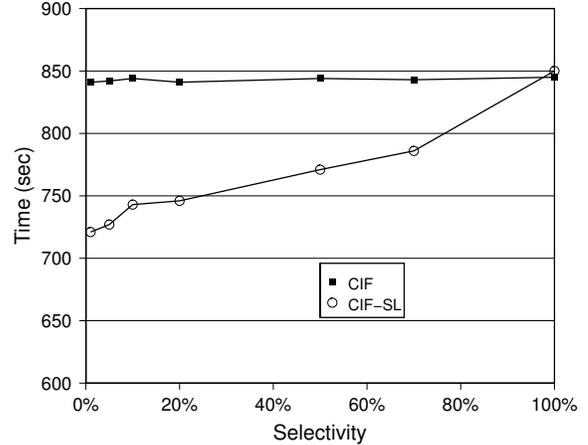

Figure 10: Benefits of lazy materialization and skip lists.

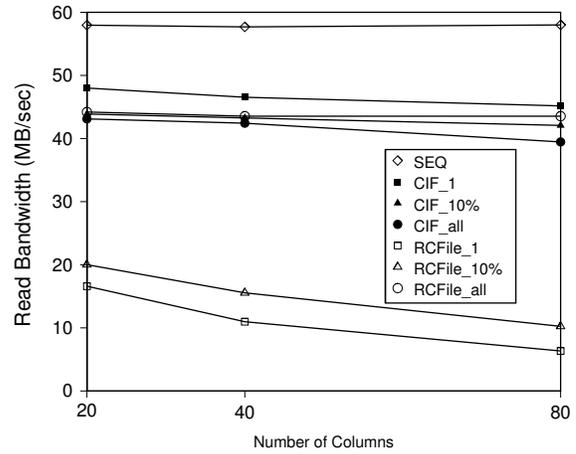

Figure 11: Comparison of CIF and RCFile as the number of columns in a record increases.

spectively. The larger row-group size clearly helped achieve better I/O elimination. However, a larger row-group size has an adverse impact on the benefits from lazy decompression as described by the authors of RCFile [20]. By eliminating this additional tuning parameter, CIF is more robust and at the same time offers better performance than RCFile.

The single integer scan was the worst case for RCFile. The performance degradation in other cases (single string, single map) was 2x-3x when using 16MB as the row-group size.

### B.3 Load Times

We measured the time taken to convert the synthetic dataset used in Section 6.2 from SEQ to various formats. These are presented in Table 2.

| Layout | Time (min) |
|--------|------------|
| CIF    | 89         |
| CIF-SL | 93         |
| RCFile | 89         |

Table 2: Load times with synthetic data.

Observe that the overhead of adding skip lists to the CIF was fairly minor. We expect this cost to be representative. The additional overhead comes from the fact HDFS exposes an append-only API. While writing the output of a job, one can only append to the file. It is not possible to go back and alter any values. As a result, building skip lists requires double buffering the data so one can actually calculate the number of bytes for each skip pointer before writing the data to disk. With the current load algorithm, the largest skip is limited by the size of the main memory. Another observation from Table 2 is that converting to uncompressed RCFiles takes approximately the same amount of time as CIF. We do not expect the load utilities that convert data to CIF to be any worse than those that convert data to RCFile.

### B.4 Varying Selectivity

Working with the same dataset as in Section 6.2, we measured the benefits of skip lists and lazy deserialization as the selectivity of the predicate in the map function was varied. We measured the time taken to aggregate the value in the map-typed column under a given key for all the records where the string column satisfied a given pattern. We varied the selectivity of the predicate and measured the running time of the job. We compared the running time of CIF vs CIF-SL. The results are shown in Figure 10.

The figure shows that for highly selective queries, CIF-SL provides more savings by eliminating unnecessary deserialization and object creation. As the selectivity gets closer to 100% CIF-SL converges to the performance of CIF. The overhead for CIF-SL with respect to CIF at 100% selectivity is minor. The performance benefits of CIF-SL depend on the complex type, and the associated cost of deserializing it.

### B.5 Varying Record Size

In this experiment, we compared the performance of CIF and RCFile as the number of columns in a record increases.

We generated three datasets with 20, 40, and 80 columns per record. Each column contained a random string of length 30. In each case, the total data size was approximately 60GB. We conducted three scan tests where we projected 1 column, 10% of the columns, or all the columns of the dataset. For the RCFile, we used 16MB as the row-group size. Figure 11 reports the read bandwidth measured for each of the scan tests.

The figure shows that when projecting a small number of columns, CIF performs better than RCFile in all cases. The overhead of CIF over SEQ when scanning all the columns of a dataset is greater as the number of columns in the dataset increases. This is consistent with previous research on the overheads of column oriented storage [25]. Another interesting observation is that as the number of columns in the dataset increases, the read bandwidth for reading a single column decreases for RCFile, while it remains relatively stable for CIF. This is because in a wider row, the amount of data corresponding to a single column in a row group (16MB) decreases, and consequently the overheads associated with processing a row-group are amortized over fewer records.